\documentclass[reprint,superscriptaddress,aps]{revtex4-1}
\usepackage{amsmath}
\usepackage{amssymb}
\usepackage{bm}
\usepackage{braket}
\usepackage{color}
\usepackage[varg]{txfonts}
\usepackage{here}
\usepackage{varwidth}
\usepackage{dcolumn}
\usepackage[breaklinks,colorlinks=true,linkcolor=blue,urlcolor=cyan,citecolor=blue]{hyperref}
\usepackage{graphicx}

\begin{document}

\title{XRD study of the magnetization plateau above 40~T in the frustrated helimagnet CuGaCr$_{4}$S$_{8}$}

\author{M.~Gen}
\email{gen@issp.u-tokyo.ac.jp}
\affiliation{Institute for Solid State Physics, University of Tokyo, Kashiwa 277-8581, Japan}
\affiliation{RIKEN Center for Emergent Matter Science (CEMS), Wako 351-0198, Japan}

\author{K.~Noda}
\affiliation{Department of Engineering Science, University of Electro-Communications, Chofu, Tokyo 182-8585, Japan}

\author{K.~Shimbori}
\affiliation{Department of Engineering Science, University of Electro-Communications, Chofu, Tokyo 182-8585, Japan}

\author{T.~Tanaka}
\affiliation{Department of Engineering Science, University of Electro-Communications, Chofu, Tokyo 182-8585, Japan}

\author{D.~Bhoi}
\affiliation{Department of Engineering Science, University of Electro-Communications, Chofu, Tokyo 182-8585, Japan}

\author{K.~Seki}
\affiliation{Department of Engineering Science, University of Electro-Communications, Chofu, Tokyo 182-8585, Japan}

\author{H.~Kobayashi}
\affiliation{Department of Engineering Science, University of Electro-Communications, Chofu, Tokyo 182-8585, Japan}

\author{K.~Gautam}
\affiliation{RIKEN Center for Emergent Matter Science (CEMS), Wako 351-0198, Japan}

\author{M.~Akaki}
\affiliation{Institute for Materials Research, Tohoku University, Sendai, Miyagi 980-8577, Japan}

\author{Y.~Ishii}
\affiliation{Institute for Solid State Physics, University of Tokyo, Kashiwa 277-8581, Japan}

\author{Y.~H.~Matsuda}
\affiliation{Institute for Solid State Physics, University of Tokyo, Kashiwa 277-8581, Japan}

\author{Y.~Kubota}
\affiliation{RIKEN SPring-8 Center, 1-1-1 Kouto, Sayo, Hyogo 679-5148, Japan}

\author{Y.~Inubushi}
\affiliation{RIKEN SPring-8 Center, 1-1-1 Kouto, Sayo, Hyogo 679-5148, Japan}
\affiliation{Japan Synchrotron Radiation Research Institute, 1-1-1 Kouto, Sayo, Hyogo 679-5198, Japan}

\author{M.~Yabashi}
\affiliation{RIKEN SPring-8 Center, 1-1-1 Kouto, Sayo, Hyogo 679-5148, Japan}
\affiliation{Japan Synchrotron Radiation Research Institute, 1-1-1 Kouto, Sayo, Hyogo 679-5198, Japan}

\author{Y.~Kohama}
\affiliation{Institute for Solid State Physics, University of Tokyo, Kashiwa 277-8581, Japan}

\author{T.~Arima}
\affiliation{RIKEN Center for Emergent Matter Science (CEMS), Wako 351-0198, Japan}

\author{A.~Ikeda}
\affiliation{Department of Engineering Science, University of Electro-Communications, Chofu, Tokyo 182-8585, Japan}

\begin{abstract}

CuGaCr$_{4}$S$_{8}$, which contains a chromium breathing pyrochlore network, exhibits diverse magnetic phases, including an incommensurate helical state below 31~K and a 1/2-magnetization plateau above 40~T, owing to the interplay between magnetic frustration and spin-lattice coupling.
Here, we perform a single-shot powder x-ray diffraction experiment on CuGaCr$_{4}$S$_{8}$ in a pulsed high magnetic field of 55~T, revealing an orthorhombic-to-cubic (or pseudocubic) structural transition upon entering the 1/2-magnetization plateau phase at low temperatures.
This observation suggests the emergence of a commensurate ferrimagnetic order, where a 3-up--1-down spin configuration is realized in each small tetrahedron, and the all-up or all-down in each large tetrahedron.
We propose two types of 16-sublattice magnetic structures, which are degenerate within exchange interactions between the first, second, and third nearest neighbors.

\end{abstract}

\date{\today}
\maketitle

\section{\label{Sec1}Introduction}

In frustrated magnets with competing magnetic interactions, the application of a magnetic field can induce successive phase transitions, accompanied by changes in the magnetic unit cell.
Neutron scattering/diffraction experiments under magnetic fields are essential for understanding these phenomena, although the required field strengths often exceed the capability of conventional superconducting magnets.
Over the last three decades, notable technical advancements have been made in neutron experiments combined with a nondestructive pulsed magnet \cite{1991_Noj, 2018_Duc}, contributing to the identification of high-field phases in various magnetic systems, such as the triangular lattice magnet CuFeO$_{2}$ \cite{1999_Mit, 2024_Nak}, the Shastry-Sutherland lattice magnet TbB$_{4}$ \cite{2009_Yos, 2022_Qur}, and several multiferroic compounds \cite{2011_Noj, 2020_Fog, 2024_Hol}.
X-ray diffraction (XRD) can be another useful tool for gaining insights into magnetic structures \cite{2013_Mat}.
A representative example is CdCr$_{2}$O$_{4}$ \cite{2005_Chu}, known as a prototypical Heisenberg antiferromagnet on a pyrochlore lattice.
One characteristic feature of CdCr$_{2}$O$_{4}$ is the appearance of a 1/2-magnetization plateau in a broad field range between 28 and 58~T \cite{2008_Koj}, suggesting that a 3-up--1-down ferrimagnetic state is stabilized by the spin-lattice coupling \cite{2004_Pen, 2006_Ber}.
Single-crystal XRD experiments in pulsed magnetic fields by Inami {\it et al.} \cite{2006_Ina} first revealed that the ferrimagnetic state retains cubic symmetry, rather than the alternative possibility of rhombohedral symmetry.
This finding was later corroborated by neutron scattering experiments by Matsuda {\it et al.} \cite{2010_Mat}.

\begin{figure}[t]
\centering
\includegraphics[width=0.75\linewidth]{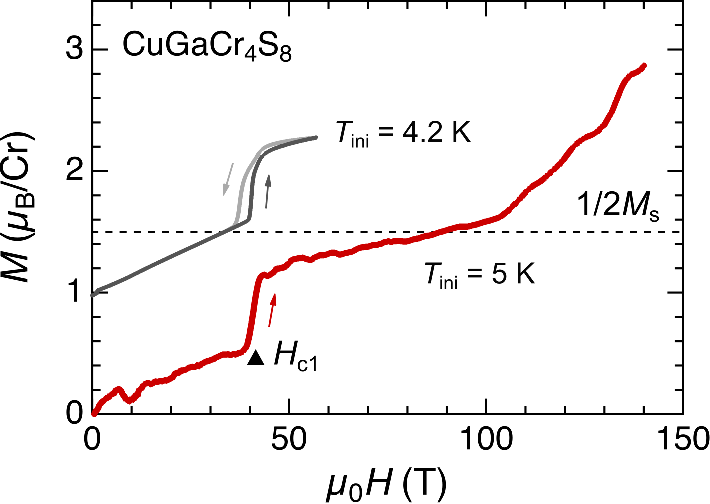}
\caption{Magnetization curves of CuGaCr$_{4}$S$_{8}$ at an initial temperature of $T_{\rm ini} = 5$~K (4.2~K), measured using the single-turn coil system (red) [using a nondestructive pulsed magnet (gray)] at the Institute for Solid State Physics, University of Tokyo, Japan. The data are reproduced from Ref.~\cite{2023_Gen}.}
\label{Fig1}
\end{figure}

Interestingly, a robust 1/2-magnetization plateau also appears in a {\it breathing} pyrochlore magnet CuGaCr$_{4}$S$_{8}$ \cite{2023_Gen, 2024_Gen}, as shown in Fig.~\ref{Fig1}.
CuGaCr$_{4}$S$_{8}$ possesses a spinel structure similar to that of CdCr$_{2}$O$_{4}$, while the crystallographic ordering of the nonmagnetic Cu$^{+}$ and Ga$^{3+}$ ions, resembling a zinc-blende-type arrangement, leads to the loss of inversion symmetry, resulting in the $F{\overline 4}3m$ space group (No.~216).
The Cr$^{3+}$ ions, with orbital-quenched $S = 3/2$ moments, form a breathing pyrochlore lattice, where the nearest-neighbor exchange interactions within the small and large tetrahedra are antiferromagnetic (AFM) and ferromagnetic (FM), respectively: $J/k_{\rm B} = 9.8(7)$~K and $J'/k_{\rm B} = -11.4(6)$~K, where $k_{\rm B}$ is the Boltzmann's constant \cite{2023_Gen}.
In zero magnetic field, a magnetic transition occurs at $T_{\rm N} = 31$~K, accompanied by a structural transition to the orthorhombic $Imm2$ space group (No.~44) \cite{2023_Gen, 2024_Gen}.
A powder neutron scattering revealed an incommensurate magnetic propagation vector of ${\mathbf Q} = (0.31, 0.5, 0)$ in the orthorhombic setting, where a cycloidal-type order was proposed as a candidate magnetic structure \cite{2024_Gen, 1976_Wil}.
Upon applying a magnetic field, CuGaCr$_{4}$S$_{8}$ undergoes a metamagnetic transition with hysteresis at $\mu_{0}H_{\rm c1} = 40$~T, followed by a plateau-like behavior at approximately half the saturation magnetization $M_{\rm s}$, persisting up to 105~T \cite{2023_Gen} (Fig.~\ref{Fig1}).
This phenomenon is reminiscent of the emergence of a 3-up--1-down state, although direct microscopic evidence remains elusive due to the formidable challenges of conducting neutron scattering experiments above 40~T.
Given the distinct spin Hamiltonians of CdCr$_{2}$O$_{4}$ and CuGaCr$_{4}$S$_{8}$, a compelling question arises regarding how the symmetry of the ferrimagnetic state in CuGaCr$_{4}$S$_{8}$ differs from that in CdCr$_{2}$O$_{4}$.

The authors of this study have recently pioneered x-ray diffractometry under ultrahigh magnetic fields exceeding 100~T by combining a destructive pulsed magnet with an x-ray free-electron laser (XFEL) \cite{2022_Ike, 2025_Ike}.
This achievement was made possible through the development of portable pulsed high-field generators utilizing the single-turn coil (STC) technique.
The first apparatus, PINK-01, equipped with a 10.4~$\mu$F main capacitor rated at 30 kV, successfully generated magnetic fields up to 77~T \cite{2022_Ike}.
Its upgraded successor, PINK-02, features two capacitors connected in parallel (20.8~$\mu$F, 30~kV) and is capable of generating magnetic fields as high as 120~T \cite{2025_Ike}.
These PINK systems have been installed on the XFEL beamline at SACLA, Japan, which provides a high-brightness, short-pulse x-ray source \cite{2012_Tan, 2019_Ino}.
The XFEL's pulse duration, on the order of several tens of femtoseconds, is sufficiently short compared to the microsecond-scale pulsed field generation.
This enables the acquisition of XRD data at a specific magnetic field in a single-shot experiment.
Additionally, we have developed a liquid-$^{4}$He-flow cryogenic system that can operate within the narrow field-generation space ($< \phi 5$~mm) of the STC \cite{2025_Ike}.

In this paper, we report a structural study on the 1/2-magnetization plateau phase of CuGaCr$_{4}$S$_{8}$ through our newly-developed single-shot XRD technique in pulsed magnetic fields up to 55~T.
We observe an orthorhombic-to-cubic (or pseudocubic) structural transition associated with the metamagnetic transition.
The identified cubic lattice symmetry is compatible with the 3-up--1-down state expected in the breathing pyrochlore magnet with AFM $J$ and FM $J'$.
Based on the microscopic spin Hamiltonian taking into account further-neighbor exchange interactions, we discuss the possible arrangement of the up- and down-spins in the ferrimagnetic state in CuGaCr$_{4}$S$_{8}$, which should be distinct from that realized in CdCr$_{2}$O$_{4}$ \cite{2010_Mat}.

\vspace{-0.2cm}
\section{\label{Sec2}Methods}

\vspace{-0.2cm}
Polycrystalline samples of CuGaCr$_{4}$S$_{8}$ were synthesized by the solid-state reaction method, as described in Ref.~\cite{2023_Gen}.
For the XRD experiments, the powdered samples were mixed with Stycast1266 in a 1:1 mass ratio and then sandwiched between two layers of Kapton sheet (20~$\mu$m thickness) to form a film.
For the magnetocaloric effect (MCE) measurements, we prepared a disk-shaped sintered sample with a dimension of $6 \times 6 \times 1$~mm$^{3}$.

\begin{figure}[t]
\centering
\includegraphics[width=\linewidth]{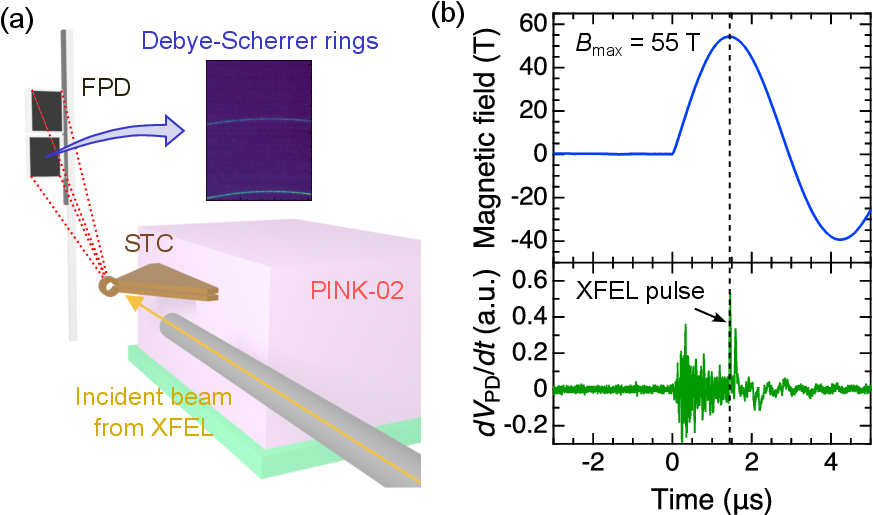}
\caption{(a) Schematic of the experimental setup for powder XRD using the STC technique at the PINK-02 system on the XFEL beamline at SACLA. A portion of the Debye-Scherrer rings from the sample, located inside the STC, is recorded using a flat panel detector (FPD). (b) The top panel shows the waveform of pulsed magnetic fields generated by PINK-02 with the STC of $\phi 5$~mm at a charging voltage of 20~kV. The bottom panel shows the timing of the XFEL pulse, detected using a photodiode. The discharge timing is synchronized with the XFEL pulse to enable the detection of a single-shot XRD at the maximum magnetic field of 55~T.}
\label{Fig2}
\end{figure}

Single-shot XRD experiments in pulsed high-magnetic fields were performed at beamline BL3EH2 of the XFEL facility SACLA, Japan \cite{2012_Tan}.
Figure~\ref{Fig2}(a) schematically illustrates the experimental setup for the powder XRD.
A seeded beam \cite{2019_Ino} was introduced at a photon energy of $E = 10.3714$~keV with energy bandwidth of $\sim$2~eV.
For high-magnetic field generation, we used the portable pulsed power system PINK-02 \cite{2022_Ike, 2025_Ike}.
By utilizing a STC with an inner diameter of 5~mm and applying a 20~kV discharge voltage, a pulsed magnetic field of up to $55 \pm 3$~T with a duration of $\sim$3~$\mu$s was generated, as shown in Fig.~\ref{Fig2}(b).
A portion of the Debye-Scherrer rings from the sample was recorded with a forward scattering geometry using a flat panel detector (FPD) with the detection window of $2\theta = 22.5^{\circ}$--$40.5^{\circ}$.
The full width at half maximum (FWHM) of the Bragg peaks was $2\theta \approx 0.05^{\circ}$.
Within the detection range up to $2\theta = 40^{\circ}$, lattice distortions of at least 0.1~\% can be observed as peak splitting or broadening in the XRD data.
For cooling the sample, we used a handmade Liquid-$^{4}$He-flow cryostat made of glass epoxy (G-10) with an outer diameter of 2.6~mm \cite{2025_Ike}.

The MCE up to 55~T was measured under the adiabatic condition using a nondestructive pulsed magnet ($\sim$36 ms duration) at the Institute for Solid State Physics, University of Tokyo, Japan \cite{2013_Kih}.
A sensitive Au$_{16}$Ge$_{84}$ film thermometer was sputtered on the surface of the sintered sample, with temperature calibration performed using a commercial RuO$_{2}$ thermometer.

\begin{figure*}[t]
\centering
\includegraphics[width=\linewidth]{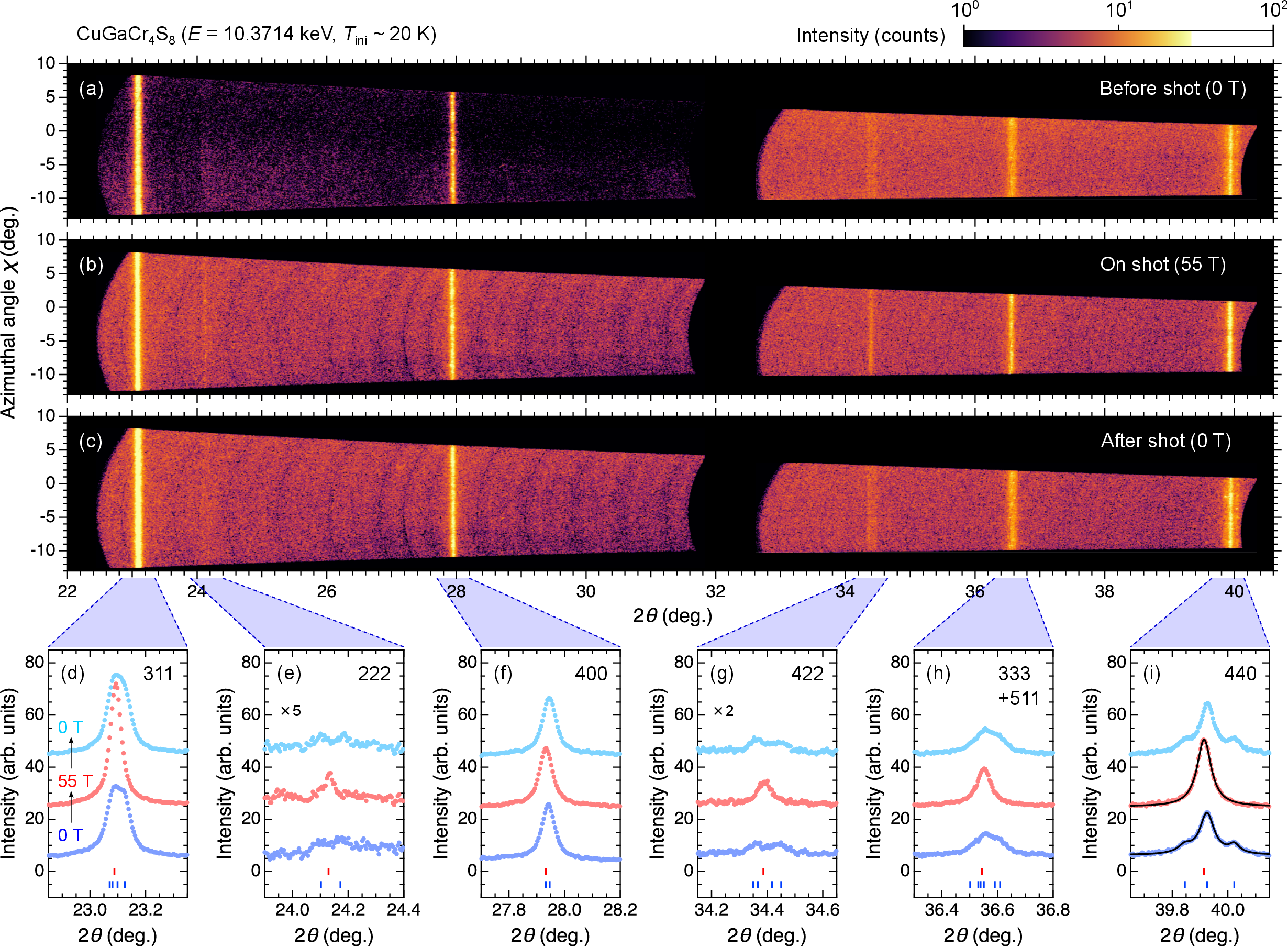}
\caption{[(a)--(c)] Images of a series of single-shot XRD from polycrystalline samples of CuGaCr$_{4}$S$_{8}$ measured at an initial temperature of $T_{\rm ini} \sim 20$~K: (a) at 0~T just before the pulsed-field shot, (b) at 55~T, and (c) at 0~T just after the pulsed-field shot. A seeded XFEL beam with $E = 10.3714$~keV was used. The data for $2\theta = 31.8^{\circ}$--$32.7^{\circ}$ are outside the detection range. The intensity is plotted on a logarithmic scale, as shown in the color scale at the top right side of panel~(a). [(d)--(i)] One-dimensional XRD profiles obtained by integrating the XRD image data shown in panels (a)--(c) in the azimuthal direction, focusing on the (d) 311, (e) 222, (f) 400, (g) 422, (h) 333 and 511, and (i) 440 Bragg reflections in the cubic setting. The blue and red bars indicate the calculated Bragg peak positions at 0~T before the shot and at 55~T, respectively, based on the (multi-)Lorentzian fit on the 440 Bragg peak(s), as shown by black solid lines in panel (i).}
\label{Fig3}
\end{figure*}

\vspace{-0.2cm}
\section{\label{Sec3}Results}

\vspace{-0.2cm}
\subsection{\label{Sec3-1}Single-shot XRD at 55~T}

\vspace{-0.2cm}
Figures~\ref{Fig3}(a)--\ref{Fig3}(c) show a series of single-shot XRD images obtained consecutively at 100~ms intervals: before the shot (0~T) [Fig.~\ref{Fig3}(a)], at the maximum field (55~T) [Fig.~\ref{Fig3}(b)], and after the shot (0~T) [Fig.~\ref{Fig3}(c)].
Although we did not perform an {\it in situ} measurement of the sample temperature due to the limited sample space, the sample temperature just before the field generation was likely approximately 20~K (much lower than $T_{\rm N} = 31$~K) based on the observed Bragg peak splitting, as discussed below.
For each data, all the expected Bragg peaks are clearly observed with a low background level, except for 331 at $2\theta \approx 30.5^{\circ}$ and 420 at $2\theta \approx 34.4^{\circ}$ with very weak intensity \cite{2023_Gen, 2024_Gen} (indices are based on the original cubic cell throughout the manuscript).
By integrating the XRD image data in the azimuthal direction, we obtain one-dimensional XRD patterns, as shown in Figs.~\ref{Fig3}(d)--\ref{Fig3}(i).

At zero field both before and after the shot, peak splitting or broadening is evident for many Bragg reflections, as reported in Ref.~\cite{2024_Gen}; the 222 reflection splits into two peaks, and the 440 reflection splits into three peaks, while the 400 reflection shows no clear splitting.
The observed peak splitting patterns are consistent with orthorhombic symmetry with the $Imm2$ space group, characterized by a rhombic lattice distortion.
By performing a triple Lorentzian fit on the 440 reflection observed before the shot [bottom black line in Fig.~\ref{Fig3}(i)], the lattice parameters of the original $F$ lattice are estimated to $a = 9.90136$~$\AA$, $c = 9.90579$~$\AA$, $\gamma = 89.756^{\circ}$.
The calculated peak positions based on these lattice parameters are displayed by blue bars for all the indices [Figs.~\ref{Fig3}(d)--\ref{Fig3}(i)], which agree with the experimental peak positions.
According to Ref.~\cite{2024_Gen}, the rhombic distortion is $\gamma = 89.743^{\circ}$ at 4~K, $\gamma = 89.748^{\circ}$ at 20~K, and $\gamma = 89.784^{\circ}$ at 30~K.
We hence infer that the initial sample temperature in this measurement would be approximately 20~K.
It should be noted that strain induced by the adhesive (Stycast1266) may reduce the distortion in the sample, suggesting that the actual sample temperature might be lower than 20~K.

Remarkably, all the Bragg reflections change to a single peak at 55~T.
This suggests that the metamagnetic transition at $H_{\rm c1} = 40$~T \cite{2023_Gen} accompanies a structural transition from orthorhombic to cubic (or pseudocubic) symmetry.
From the Lorentzian fit on the 440 reflection [middle black line in Fig.~\ref{Fig3}(i)], the lattice parameter at 55~T is estimated to $a = 9.90589~\AA$, yielding the volume expansion of $\Delta V/V \approx 9.4 \times 10^{-4}$.
This is consistent with the linear magnetostriction of $\Delta L/L \approx 3 \times 10^{-4}$ across $H_{\rm c1}$ observed by the fiber-Bragg-grating method \cite{2023_Gen}, given that $\Delta V/V \approx 3(\Delta L/L)$.

\vspace{-0.2cm}
\subsection{\label{Sec3-2}Magnetocaloric effect}

\vspace{-0.2cm}
As the field-induced phase in CuGaCr$_{4}$S$_{8}$ is characterized by a 1/2-magnetization plateau \cite{2023_Gen}, the emergence of a 3-up--1-down state is the most plausible scenario, similar to the case in conventional Cr spinel oxides \cite{2008_Koj, 2010_Mat, 2006_Ued, 2007_Mat}.
However, before drawing this conclusion, it is necessary to rule out the possibility of a paramagnetic state appearing at 55~T due to sample heating caused by a hysteresis loss and the magnetocaloric effect (MCE).
Since the field duration is only approximately 3~$\mu$s in our high-field XRD experiment [Fig.~\ref{Fig2}(b)], the measurement conditions should be adiabatic, where the sum of the lattice entropy and the magnetic entropy is preserved.
Therefore, the sample temperature should change in association with the change in the magnetic entropy upon the application of a magnetic field.

\begin{figure}[t]
\centering
\includegraphics[width=0.8\linewidth]{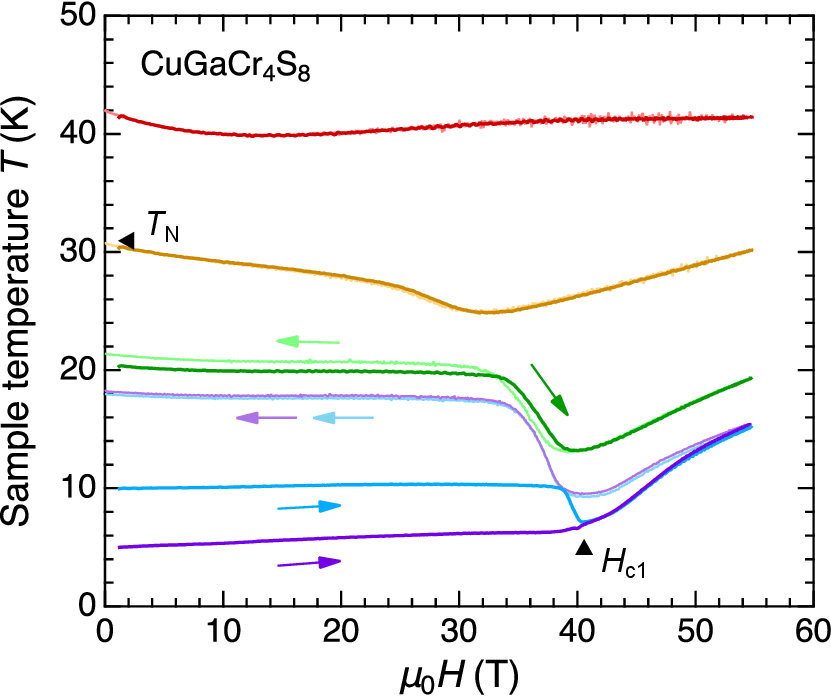}
\caption{Magnetocaloric effect in polycrystalline samples of CuGaCr$_{4}$S$_{8}$ under the adiabatic condition measured at various initial temperatures. The thick (thin) lines correspond to the data in the field-increasing (decreasing) process.}
\label{Fig4}
\end{figure}

Figure~\ref{Fig4} shows the field dependence of the sample temperature $T(H)$ up to 55~T measured at various initial temperatures ($T_{\rm ini}$) under adiabatic conditions.
The cooling of the sample in the vicinity of $H_{\rm c1}$ (except for $T_{\rm ini} = 5$~K) indicates the accumulation of the magnetic entropy near the phase boundary, and the subsequent heating above $H_{\rm c1}$ suggests the opening of the spin gap in the field-induced phase.
These MCE trends are qualitatively similar to those observed in other Cr spinels \cite{2022_Kim, 2022_Gen}.
We note that, for both $T_{\rm ini} = 5$~K and 10~K, the sample temperature eventually rises to 18~K at 0~T after the field-down sweep, indicating a significant hysteresis loss associated with the first-order metamagnetic transition at $H_{\rm c1}$.
Importantly, the sample temperature never exceeds 30~K at 55~T as long as the initial sample temperature is below 30~K.
We hence conclude that the cubic (or pseudocubic) phase observed in the high-field XRD experiment is not a paramagnetic state.

\begin{figure}[t]
\centering
\includegraphics[width=0.95\linewidth]{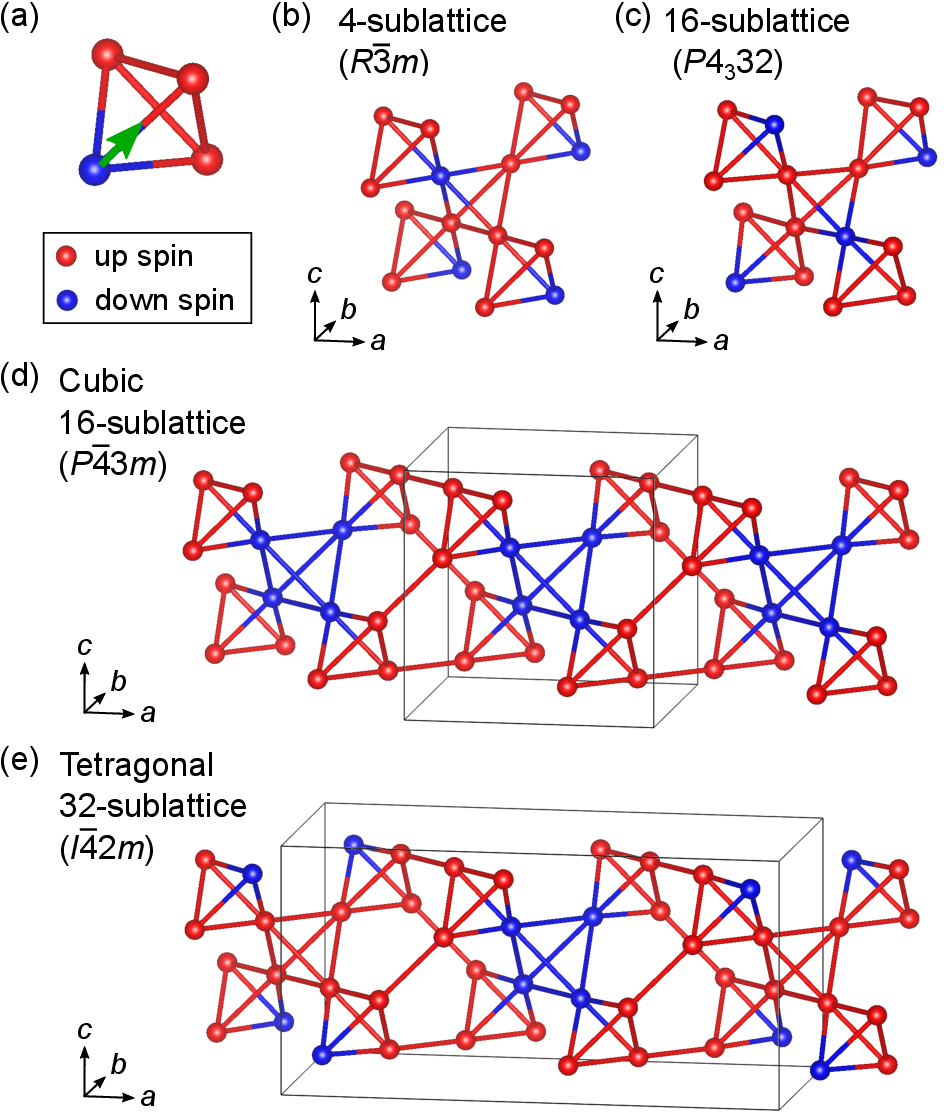}
\caption{(a) Schematic illustration of the compressive trigonal distortion in a tetrahedron with a 3-up--1-down spin configuration. The down-spin site shifts towards the $\langle 111 \rangle$ axis (or its equivalents), i.e., the center of gravity of the triangle formed by the three up-spin sites, as indicated by the green arrow, resulting in three longer FM bonds (red) and three shorter AFM bonds (red-blue). [(b)(c)] Possible two types of 3-up--1-down spin configurations in the pyrochlore antiferromagnet: (b) 4-sublattice structure with the $R{\overline 3}m$ symmetry and (c) 16-sublattice structure with the $P4_{3}32$ (or $P4_{1x}32$) symmetry. In HgCr$_{2}$O$_{4}$ and CdCr$_{2}$O$_{4}$, the appearance of the $P4_{3}32$-type structure was experimentally confirmed \cite{2006_Ina, 2010_Mat, 2007_Mat}. [(d)(e)] Possible two types of 3-up--1-down spin configurations in the breathing pyrochlore magnet with AFM $J$ and FM $J'$: (d) Cubic 16-sublattice structure with the $P{\overline 4}3m$ symmetry and (e) Tetragonal 16-sublattice structure with the $I{\overline 4}2m$ symmetry. The magnetic propagation vectors are ${\mathbf Q} = (1, 0, 0)$ and (1/2, 0, 0), respectively. Black boxes represent a crystallographic unit cell. Crystal structures are visualized using the VESTA software \cite{2011_Mom}.}
\label{Fig5}
\end{figure}

\vspace{-0.2cm}
\section{\label{Sec4}Discussion}

In the pyrochlore Heisenberg antiferromagnet, geometrical frustration leads to macroscopic degeneracy in the ground state.
One of the primary perturbations that can lift the degeneracy is spin-lattice coupling, which acts as a biquadratic exchange interaction between adjacent spins, favoring collinear spin configuration \cite{2004_Pen, 2006_Ber, 2010_Sha, 2016_Aoy, 2021_Aoy}.
This is evident in the stabilization of the 3-up--1-down state in a magnetic field, accompanied by the emergence of a 1/2-magnetization plateau \cite{2004_Pen, 2006_Ber, 2010_Sha, 2021_Aoy}.
Remarkably, similar phenomena can also occur in the breathing pyrochlore magnet with AFM $J$ and FM $J'$ in the small and large tetrahedra, respectively \cite{2020_Gen}.
In both cases, each AFM tetrahedron undergoes a compressive trigonal distortion, where the three up--down bonds shorten, and the remaining three up--up bonds elongate, as shown in Fig.~\ref{Fig5}(a).
However, it remains nontrivial how such a 3-up--1-down tetrahedron can be arranged across the entire (breathing) pyrochlore network.

Let us first review the discussions that have been made thus far on the conventional pyrochlore antiferromagnet case \cite{2006_Ber, 2006_Ina, 2010_Mat, 2007_Mat, 2010_Sha, 2021_Aoy}.
There are two possible spin arrangements where the spins in each tetrahedron adopt a 3-up--1-down configuration: a 4-sublattice structure with rhombohedral $R{\overline 3}m$ symmetry and a 16-sublattice structure with cubic $P4_{3}32$ (or $P4_{1}32$) symmetry, as shown in Figs.~\ref{Fig5}(b) and \ref{Fig5}(c), respectively.
The simplest theoretical framework to microscopically describe the spin-lattice coupling is the bond-phonon model proposed by Penc {\it et al}. \cite{2004_Pen}, which assumes independent displacements of each bond.
The effective spin Hamiltonian, including the Zeeman term, is given by
\begin{equation}
\label{Eq1}
{\mathcal{H}}_{\rm BP} = J\sum_{\langle ij\rangle}\left[{\mathbf S}_{i} \cdot {\mathbf S}_{j} - b({\mathbf S}_{i} \cdot {\mathbf S}_{j})^{2} \right] - {\mathbf h} \cdot \sum_{i}{\mathbf S}_{i},
\end{equation}
where ${\mathbf S}_{i}$ is a classical three-dimensional vector spin with unit length, $J$ is the nearest-neighbor AFM exchange interaction between sites $i$ and $j$, $b$ represents the strength of the spin-lattice coupling, and ${\mathbf h}$ is the external magnetic field.
Under the Hamiltonian in Eq.~\eqref{Eq1}, the 4-sublattice and 16-sublattice 3-up--1-down states are energetically degenerate, and the emergence of a spin-liquid plateau state without long-range order was suggested through finite-temperature Monte Carlo simulations \cite{2010_Sha, 2021_Aoy}.
However, x-ray and neutron experiments on CdCr$_{2}$O$_{4}$ \cite{2006_Ina, 2010_Mat} and HgCr$_{2}$O$_{4}$ \cite{2007_Mat} under magnetic fields revealed the 16-sublattice structure with cubic symmetry [Fig.~\ref{Fig5}(c)].
One possible origin for this is the presence of further-neighbor exchange interactions; the second-nearest-neighbor FM (AFM) interaction or the third-nearest-neighbor AFM (FM) interaction favor the 16-sublattice (4-sublattice) structure, which is consistent with the first-principles calculations \cite{2008_Yar}.
In the Einstein site-phonon model proposed by Bergman {\it et al.} \cite{2006_Ber}, which assumes independent displacements of each site to incorporate phonons, the spin-lattice coupling gives rise to three-body quadratic terms of the form $-({\mathbf S}_{i} \cdot {\mathbf S}_{j})({\mathbf S}_{i} \cdot {\mathbf S}_{k})$, in addition to the biquadratic term already present in the bond-phonon model in Eq.~\eqref{Eq1}.
Under the 3-up--1-down constraint, these three-body terms result in the dominant third-nearest-neighbor AFM interaction $J_{3}^{\rm eff}$ compared to the second-nearest-neighbor AFM interaction $J_{2}^{\rm eff}$, where $J_{3}^{\rm eff} = 2J_{2}^{\rm eff}$, favoring the 16-sublattice order \cite{2006_Ber, 2021_Aoy}.

\begin{figure}[t]
\centering
\includegraphics[width=\linewidth]{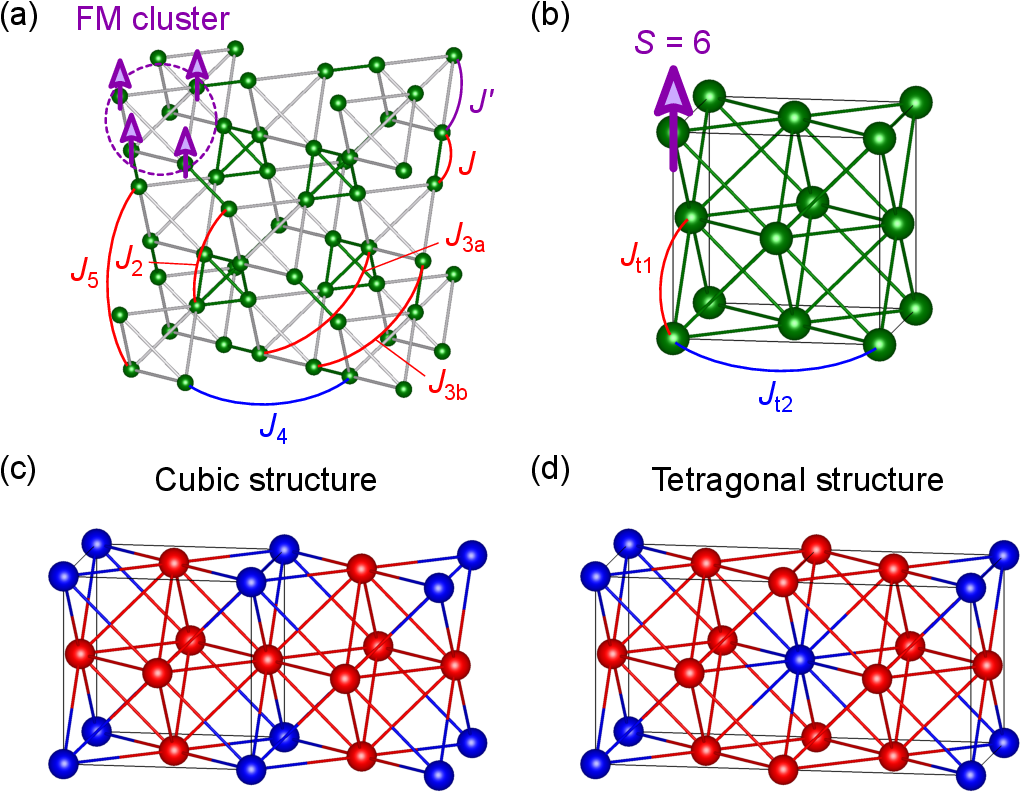}
\caption{[(a)(b)] Definition of exchange interactions up to the fifth-nearest-neighbor path on the breathing pyrochlore lattice (a). In the 3-up--1-down state of CuGaCr$_{4}$S$_{8}$, the four spins within each large tetrahedron behave as a FM cluster (i.e., all-up or all-down), resulting in an effective face-centered-cubic (FCC) lattice model consisting of an $S=6$ spin (b). Nearest-neighbor and second-nearest-neighbor exchange interactions on the effective model are $J_{\rm t1} = J + 4J_{2} + 2(J_{3a} + J_{3b} + J_{5})$ and $J_{\rm t2} = 4J_{4}$, respectively. [(c)(d)] Schematic illustrations mapping the cubic (tetragonal) 16-sublattice 3-up--1-down structure shown in Fig.~\ref{Fig5}(d) [Fig.~\ref{Fig5}(e)] onto an FCC lattice. Red and blue spheres represent all-up and all-down spin clusters, respectively. Black boxes represent a crystallographic unit cell. Crystal structures are visualized using the VESTA software \cite{2011_Mom}.}
\label{Fig6}
\end{figure}

Here, we propose the emergence of an unconventional 3-up--1-down-based ferrimagnetic state on a breathing pyrochlore lattice with AFM $J$ and FM $J'$.
For CuGaCr$_{4}$S$_{8}$, the exchange coupling parameters for the room-temperature $F{\overline 4}3m$ structure are estimated through density functional theory calculations as $J/k_{\rm B} = 9.8(7)$~K, $J'/k_{\rm B} = -11.4(6)$~K, $J_{2}/k_{\rm B} = 2.3(5)$~K, $J_{3a}/k_{\rm B} = 5.9(3)$~K, $J_{3b}/k_{\rm B} = 4.4(3)$~K, $J_{4}/k_{\rm B} = -1.0(3)$~K, and $J_{5}/k_{\rm B} = 0.4(3)$~K \cite{2023_Gen}, where each exchange path is defined as shown in Fig.~\ref{Fig6}(a).
Since the AFM $J$ and FM $J'$ are predominant compared to further-neighbor interactions, it is preferable to realize a 3-up--1-down configuration in the small tetrahedra while maintaining the all-up or all-down configuration in the large tetrahedra.
There are two possible spin arrangements that satisfy the above-mentioned constraints: a cubic structure with a magnetic propagation vector of ${\mathbf Q} = (1, 0, 0)$ and a tetragonal structure with ${\mathbf Q} = (1/2, 0, 0)$, as shown in Figs.~\ref{Fig5}(d) and \ref{Fig5}(e), respectively.
Here, we do not consider periodic mixtures of these two configurations with a long-period ${\mathbf Q}$.
In the former case, the crystallographic space group is $P{\overline 4}3m$, while in the latter, it is $I{\overline 4}2m$.
The expected atomic coordinates of the Cr sites are presented in Tables~\ref{Cubic} and \ref{Tetra} in Appendix~\ref{AppendixA}.
In both configurations, the number of the all-up tetrahedra is three times that of the all-down tetrahedra.
To gain the exchange energy in the up--down bonds, the all-down tetrahedron would expand isotropically, with each down-spin site displaced along the $\langle 111 \rangle$ direction (or its equivalents). 
As a result, the volume of the all-down tetrahedron becomes larger than that of the all-up tetrahedron while maintaining the $A_{1}$ symmetry.

We note that in the case of the tetragonal structure, no splitting of Bragg peaks is expected because the unit-cell size is simply doubled from the original cubic cell, i.e., $2a \times a \times a$.
Therefore, it is difficult to definitively determine whether the magnetic structure corresponds to ${\mathbf Q} = (1, 0, 0)$ or ${\mathbf Q} = (1/2, 0, 0)$ based solely on the changes in the peak profiles observed in our high-field XRD experiments.
In either case, superlattice reflections with integer indices, which are forbidden in the $F{\overline 4}3m$ structure, would appear.
For ${\mathbf Q} = (1/2, 0, 0)$, additional superlattice reflections with half-integer indices like (1/2 1 0) are allowed.
Observing these reflections is crucial for determining the magnetic structure; however, this was not achieved in the present powder XRD due to the insufficient intensity signal-to-noise ratio.
In Appendix~\ref{AppendixB}, we estimate the correlation between the magnitude of the local trigonal distortion of the small tetrahedron and the expected powder XRD intensities of the superlattice peaks.

Finally, we discuss which of the two 3-up--1-down states may be selected from the viewpoint of the spin Hamiltonian.
Since the four spins within each large tetrahedron behaves as a FM cluster, it is useful to consider an effective face-centered-cubic (FCC) lattice model by replacing the four spins with a single $S = 6$ spin, as shown in Fig.~\ref{Fig6}(b).
Among the exchange interactions up to the fifth nearest neighbor on the original breathing pyrochlore lattice [Fig.~\ref{Fig6}(a)], all interactions except $J'$ and $J_{4}$ are renormalized into the nearest-neighbor interaction on the FCC lattice as $J_{\rm t1} = J + 4J_{2} + 2(J_{3a} + J_{3b} + J_{5})$.
Since all of these interactions are AFM \cite{2023_Gen}, $J_{\rm t1}$ is AFM.
On the other hand, the FM $J_{4}$ contributes to the second-nearest-neighbor interaction on the FCC lattice as $J_{\rm t2} = 4J_{4}$.
Figures~\ref{Fig6}(c) and \ref{Fig6}(d) illustrate the cubic and tetragonal 16-sublattice 3-up--1-down structures shown in Figs.~\ref{Fig5}(d) and \ref{Fig5}(e) mapped onto the FCC lattice, respectively.
When only $J_{\rm t1}$ is considered, these two magnetic structures are energetically degenerate.
This degeneracy is not lifted by the spin-lattice coupling even in the framework of the site-phonon model because the phonon-mediated interactions are relevant only up to the third-nearest-neighbor path on the original breathing pyrochlore lattice \cite{2021_Aoy, 2020_Gen}.
The most primary perturbation which lifts the degeneracy would be the FM $J_{\rm t2}$, which stabilizes the cubic structure rather than the tetragonal one.
It is also noteworthy that the magnetic propagation vector of the ground state at zero field is ${\mathbf Q} = (1 - \delta, \delta, 0)$~($\delta = 0.19$) in the cubic cell \cite{2024_Gen}.
Accordingly, the appearance of the cubic 3-up--1-down state with ${\mathbf Q} = (1, 0, 0)$ would be plausible.

\vspace{-0.2cm}
\section{\label{Sec5}Summary}

\vspace{-0.2cm}
Single-shot powder XRD and complementary MCE experiments in pulsed high magnetic fields of up to 55~T reveal an orthorhombic-to-cubic (or pseudocubic) structural transition upon entering the 1/2-magnetization plateau phase above $\mu_{0}H_{\rm c1} = 40$~T in CuGaCr$_{4}$S$_{8}$.
The absence of lattice distortion is compatible with the emergence of a 3-up--1-down magnetic structure expected for the breathing pyrochlore magnet with AFM $J$ and FM $J'$.
We propose two possible spin arrangements: one is a 16-sublattice cubic structure with ${\mathbf Q} = (1, 0, 0)$, and the other is a tetragonal one with ${\mathbf Q} = (1/2, 0, 0)$.
In both cases, the crystallographic polarity is lost from the $Imm2$ structure in the zero-field helical phase \cite{2024_Gen}, which might be the origin of the magnetodielectric anomaly observed at $H_{\rm c1}$ \cite{2023_Gen}.
The present study demonstrates that XRD studies in high magnetic fields contribute significantly to the microscopic understanding of magnetic structure changes and related intriguing physical phenomena in frustrated magnets.
The high-field XRD technique can be applied up to 100-T field range, enabling the exploration of novel high-field phases across a wide range of materials.

\vspace{-0.2cm}
\section*{Acknowledgements}
This work was financially supported by MEXT LEADER program (No.~JPMXS0320210021), JST FOREST program (No.~JPMJFR222W), and the JSPS KAKENHI Grants-In-Aid for Scientific Research (No.~23K13068 and No.~24H01633).
The present experiments were performed with the approval of the Japan Synchrotron Radiation Research Institute (Proposal No.~2023A8063 and No.~2024A8033).
The authors would like to acknowledge the supports from the technical staffs of the SACLA facility.

\appendix

\vspace{-0.2cm}
\section{\label{AppendixA}Expected atomic positions of the Cr sites in the 3-up--1-down states}

\begin{figure}[t]
\centering
\includegraphics[width=\linewidth]{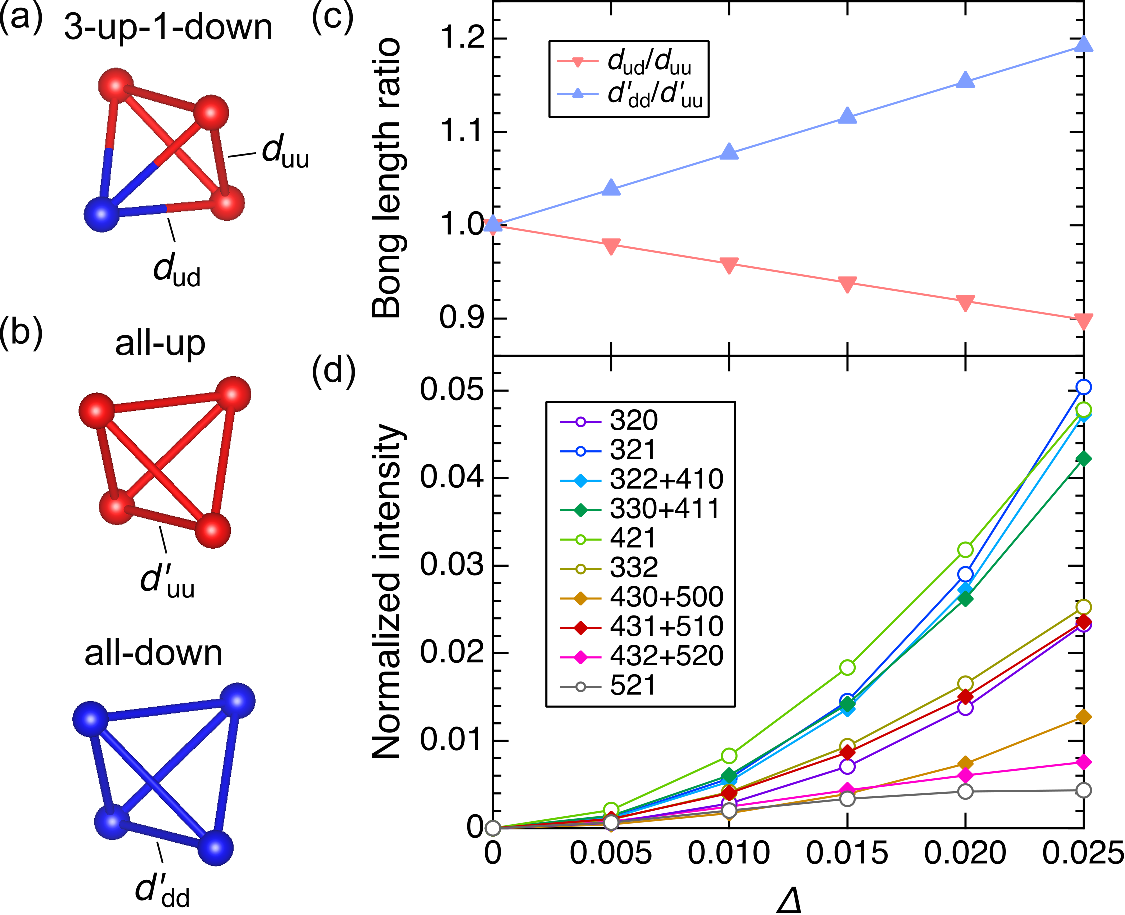}
\caption{[(a)(b)] Definition of Cr--Cr bond lengths in the 3-up--1-down state: (a) $d_{\rm uu}$ ($d_{\rm ud}$) for the up--up (up--down) bond in the small tetrahedron, and (b) $d'_{\rm uu}$ ($d'_{\rm dd}$) in the all-up (all-down) large tetrahedron. Red and blue spheres denote the up- and down-spin, respectively. (c) Ratios of Cr--Cr bond lengths, $d_{\rm ud}/d_{\rm uu}$ and $d'_{\rm dd}/d'_{\rm uu}$, as a function of $\Delta$ in the 3-up--1-down state. (d) Calculated powder XRD intensities of several superlattice peaks, normalized to the intensity of the 311 peak as a function of $\Delta$ in the 3-up--1-down state, assuming the cubic $P{\overline 4}3m$ symmetry. See the text for details.}
\label{Fig7}
\end{figure}

\vspace{-0.2cm}
Tables~\ref{Cubic} and \ref{Tetra} present the atomic coordinates of the Cr sites in the 16-sublattice 3-up--1-down structure with cubic $P{\overline 4}3m$ symmetry [Fig.~\ref{Fig5}(d)] and tetragonal $I{\overline 4}2m$ symmetry [Fig.~\ref{Fig5}(e)], respectively.
Cr1 corresponds to the down-spin sites, while Cr2 (and Cr3) correspond to the up-spin sites.
For the tetragonal structure, the lattice constants are given by $a = b = c/2$.
In the paramagnetic phase with cubic $F{\overline 4}3m$ symmetry, $\alpha \approx 0.005$ for CuGaCr$_{4}$S$_{8}$ \cite{2023_Gen, 2024_Gen}, and $\Delta = 0$.
In the 3-up--1-down state, each down-spin site shifts towards the $\langle 111 \rangle$ axis (or its equivalent), causing the up--down bond length ($d_{\rm ud}$) to become shorter than the up--up bond length ($d_{\rm uu}$) in the small tetrahedron [Fig.~\ref{Fig7}(a)].
For the large tetrahedra with the all-up or all-down spin configuration, there is no lattice distortion, although the bond length in the all-down tetrahedron ($d'_{\rm dd}$) becomes longer than that in the all-up tetrahedron ($d'_{\rm uu}$) [Fig.~\ref{Fig7}(b)].
The $\Delta$ dependence of $d_{\rm ud}/d_{\rm uu}$ and $d'_{\rm dd}/d'_{\rm uu}$ is shown in Fig.~\ref{Fig7}(c).

\begin{table}[H]
\centering
\renewcommand{\arraystretch}{1.2}
\caption{Atomic coordinates of the Cr sites in the 16-sublattice 3-up--1-down structure with cubic $P{\overline 4}3m$ symmetry.}
\begin{tabular}{ccccccc} \hline\hline
~ & ~ & $x$ & $y$ & $z$ \\ \hline
~Cr1~ & ~$4e$~ & ~$0.375 - \alpha - \Delta$~ & ~$0.375 - \alpha - \Delta$~ & ~$0.375 - \alpha - \Delta$~ \\
~Cr2~ & ~$12i$~ & ~$0.625 + \alpha$~ & ~$0.125 + \alpha$~ & ~$0.875 - \alpha$~ \\ \hline\hline
\end{tabular}
\label{Cubic}
\end{table}

\vspace{-0.5cm}
\begin{table}[H]
\centering
\renewcommand{\arraystretch}{1.2}
\caption{Atomic coordinates of the Cr sites in the 16-sublattice 3-up--1-down structure with tetragonal $I{\overline 4}2m$ symmetry. The lattice constants are $a = b = c/2$.}
\begin{tabular}{ccccccc} \hline\hline
~ & ~ & $x$ & $y$ & $z$ \\ \hline
~Cr1~ & ~$8i$~ & ~$0.125 + \alpha + \Delta$~ & ~$0.875 - \alpha - \Delta$~ & ~$(0.125 + \alpha + \Delta)/2$~ \\  
~Cr2~ & ~$8i$~ & ~$0.375 - \alpha$~ & ~$0.625 + \alpha$~ & ~$(0.125 + \alpha)/2$~ \\
~Cr3~ & ~$16j$~ & ~$0.125 + \alpha$~ & ~$0.625 + \alpha$~ & ~$(0.375 - \alpha)/2$~ \\ \hline\hline
\end{tabular}
\label{Tetra}
\end{table}

\section{\label{AppendixB}Correlation between the local trigonal distortion and powder XRD intensities of superlattice peaks}

In order to evaluate the upper limit of the magnitude of the local trigonal distortion in the 3-up--1-down state, we calculate powder XRD intensities of superlattice peaks using the VESTA software \cite{2011_Mom}.
Here, we consider only Cr atoms, as the atomic positions of Ga and S are elusive.
For simplicity, we assume the cubic $P{\overline 4}3m$ symmetry, using the structural parameters shown in Table~\ref{Cubic}.
The photon energy was set to $E = 10.3714$~keV ($\lambda = 1.1954~\AA$), the lattice parameter to $a = 9.90589~\AA$ as determined in this study, and $\alpha = 0.005$ based on the paramagnetic structure \cite{2023_Gen, 2024_Gen}.

Figure~\ref{Fig7}(d) shows the calculated intensities of several superlattice peaks within a $2\theta$ range of $22.5^{\circ}$ to $40.5^{\circ}$ as a function of $\Delta$, which is related to the degree of the trigonal distortion.
Note that the intensity is normalized to that of the 311 peak.
From the XRD data shown in Fig.~\ref{Fig3}, the minimum discernible peak intensity is approximately 5~\% of the intensity of the 311 peak.
As shown in Fig.~\ref{Fig7}(d), it is evident that $\Delta$ must be at least 0.025 for the intensity of at least one superlattice peak to reach 5~\% of the 311 peak intensity.
$\Delta = 0.025$ corresponds to the lattice distortion characterized by $d_{\rm ud}/d_{\rm uu} \approx 0.9$ and $d'_{\rm dd}/d'_{\rm uu} \approx 1.2$ [Fig.~\ref{Fig7}(c)].
The absence of superlattice peaks in the present XRD data suggests that the magnitude of the local trigonal distortion in the 3-up--1-down state is at most a few percent or less.
We note that in the low-temperature orthorhombic $Imm2$ structure at zero field, the local lattice distortion from the original regular tetrahedron is approximately 2~\% \cite{2024_Gen}.

\end{document}